# Band edge discontinuities and carrier transport in *c*-Si/porous silicon heterojunctions

Md. N. Islam[1], Sanjay K. Ram[2,3] and Satyendra Kumar[2,4]

[1]QAED-SRG, Space Applications Centre (ISRO), Ahmedabad – 380015, India
[2]Department of Physics, Indian Institute of Technology Kanpur, Kanpur- 208016, India

**Abstract**
We have prepared light emitting nanocrystalline porous silicon (PS) layers by electrochemical anodization of crystalline silicon (*c*-Si) wafer and characterized the *c*-Si/PS heterojunctions using temperature dependence of dark current-voltage (*I-V*) characteristics. The reverse bias *I-V* characteristics of *c*-Si/PS heterojunctions are found to behave like Schottky junctions where carrier transport is mainly governed by the carrier generation-recombination in the depletion region formed on the PS side. Fermi level of *c*-Si gets pinned to the defect levels at the interface resulting in $\ln(I) \propto V^{1/2}$. The barrier height in the reverse bias condition is shown to be equal to the band offset at the conduction band edges. An energy band diagram for the *c*-Si/PS heterojunction is proposed.

## 1. Introduction

The attractive optoelectronic properties of porous silicon (PS) along with the low cost fabrication possibilities have paved new ways for the development of this photoluminescent material [1,2,3]. While the optical properties of PS have been extensively studied, the electrical properties and conduction mechanisms, though of prime import, still have some contentious issues associated with them [4,5,6,7,8,9]. Differences in structural characteristics of PS and in the multijunction properties of PS based nanostructures give rise to the observed diversity of electrical properties [10]. The optoelectronic applications such as light-emitting devices [3], photodetectors and solar cells [11] and sensing devices [12,13] using PS active layer require proper understanding of electronic transport behavior of PS layers in device structures, especially the junction properties [14].

The most widely investigated device structure using PS layer is crystalline silicon (*c*-Si) /PS/metal. The electrical characteristics of metal/*c*-Si/PS/metal structures have been shown to exhibit similar rectifying features irrespective of the metal used for top contacts [5]. Therefore, transport of carriers within the PS layer thickness and across the *c*-Si/PS heterojunction governs the device characteristics. The electrical response of a *c*-Si/PS heterojunction is determined by the interplay of band edge offsets and density of defect states (DOS) in the PS layer. Most of the studies on *c*-Si/PS structures use forward current-voltage (*I-V*) characteristics to determine the transport parameters. However, high series resistance due to PS layer and high ideality factors *n* (>5) using the conventional analysis of forward *I-V* characteristics impedes a determination of reliable junction parameters in *c*-Si/PS heterojunctions [4]. In this study, we report on the reverse (*I-V*) characteristics measured as a function of temperature on *c*-Si/PS junctions. We obtained the conduction band offset at the *c*-Si/PS junction and have proposed an energy band diagram of the *c*-Si/PS heterojunction.

## 2. Sample preparation and electrical characterization

The PS layers were formed by electrochemical anodization of p-type *c*-Si (100) wafers of 6-10 $\Omega$.cm resistivity in a Teflon cell using HF (48%) and $C_2H_5OH$ (99.9%), in 1:1 proportion by volume, as electrolyte and a platinum disc as a counter electrode. For a uniform current distribution over the exposed area, an Ohmic back contact was created by thermal evaporation of Al, followed by annealing at 450° C for 1 hour, both procedures carried out in high vacuum conditions. The wafers were anodized at a constant current density of about 10 mA.cm$^{-2}$ for times varying from 90 to 120 minutes under white light illumination, resulting in 30–50 $\mu$m thick PS layers. Samples were rinsed in deionized water followed by methanol and subsequently soaked in propanol for a few minutes to minimize the structural damage during drying. Photoluminescence measurements and Raman spectroscopy using a micro-probe were used to characterize the samples [15,16]. Careful data analysis revealed a nanocrystalline Si structure having a normal crystallite size distribution [15,17].

---

[3] Corresponding author. E-mail address: skram@iitk.ac.in, sanjayk.ram@gmail.com
[4] satyen@iitk.ac.in

For electrical measurements, Al pads of $\approx$ 2 mm diameter were immediately thermally evaporated on top of the porous layers in glancing geometry at an angle of 30° between molecular beam and the sample. This precaution prevents shorting of contact between evaporated metal and the silicon skeleton (especially for thick PS layers). In order to make an intimate contact between PS and Al, samples were annealed at $\approx$ 200 °C for 45 min. All electrical measurements were carried out in a closed-cycle helium cryostat under dark conditions. Care was taken to avoid any light or thermal induced degradation [18].

## 3. Results

In our device structure of Al/$c$-Si/PS/Al, measured *I-V* characteristics may be governed by either $c$-Si/PS heterojunction or PS/Al interface, or both. In earlier studies, we found PS/Al junctions to exhibit an Ohmic or quasi-linear *I-V* characteristics [19]. Several research groups have also reported nearly Ohmic nature of PS/Al junctions [20,21,22,23,24]. Therefore, we shall focus our attention on the transport across $c$-Si/PS heterojunctions.

Figure 1 shows a set of *I-V* characteristics measured on Al/$c$-Si/PS/Al structures at different temperatures from 150 K to 300 K. The characteristics show rectifying behavior and are found to be reversible and reproducible over the temperature range studied. In order to take steady state measurement at lower temperatures (<200 K), each data point was taken only when three successive readings taken at 10-second intervals remained within a range of 5 % variations. The Al/$c$-Si/PS/Al device structures exhibited higher current conduction when the top Al contact was biased negatively with respect to the $c$-Si. However, positive voltages are shown as forward bias in figure 1 as a matter of convention. The reverse current increases slowly with increasing reverse bias. The lack of reverse saturation is reminiscent of Schottky diodes where barrier height depends on the electric field in the depletion region. Generally, for highly resistive semiconducting material or small junction barrier height, the diode equation is conventionally written in the form [25]:

$$I = I_0 \exp\left(\frac{q(V - IR_s)}{nkT}\right)\left[1 - \exp\left(\frac{-q(V - IR_s)}{kT}\right)\right] \quad (1)$$

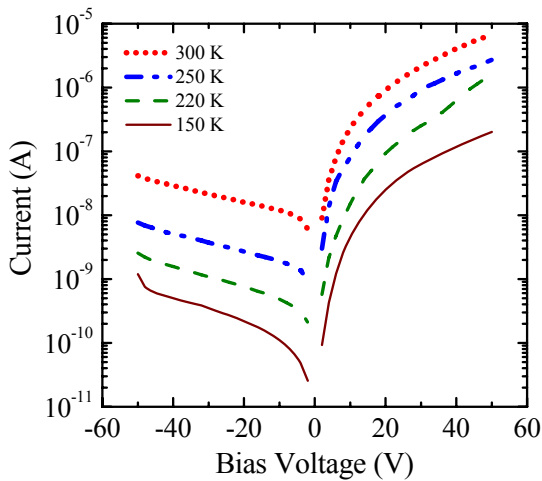

Figure 1. *I-V* characteristics of a Al/PS/$c$-Si/Al structure measured at different temperatures.

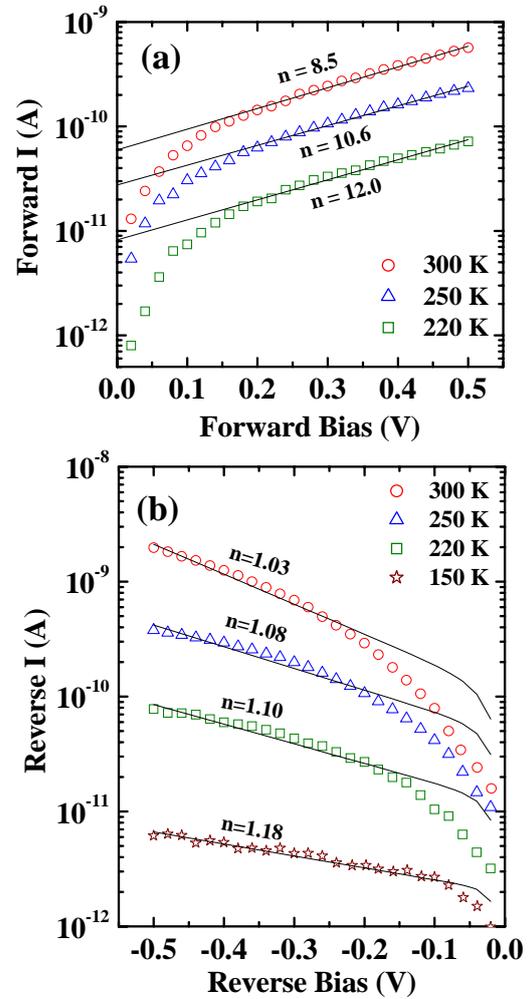

Figure 2. Fitting of forward (a) and reverse (b) *I-V* characteristics to Eq.1 at different temperatures.

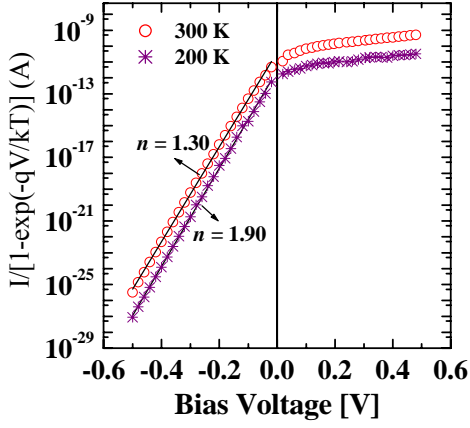 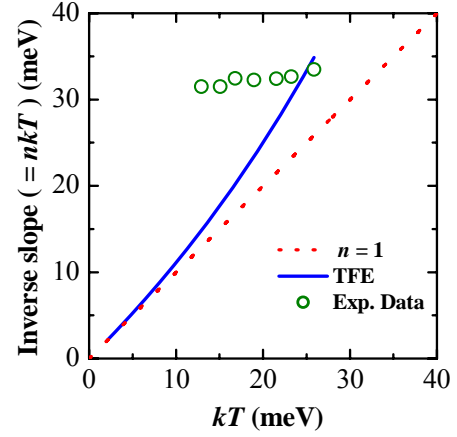

Figure 3. Semi logarithmic plots of $I/\{1-\exp(-qV/kT)\}$ as a function of bias voltage at two different temperatures (see Eq. 3 for details).

Figure 4. Plot of inverse slope ($nkT$) vs $kT$ showing different types of temperature dependence of ideality factor: Thermionic emission model (doted line), TFE model (solid line) and our data (circles).

where $I_0$ is the saturation current, $R_S$ is the series resistance of the device structure, $n$ is the ideality factor, and $k$ is the Boltzmann constant.

Figure 2 shows the forward and reverse $I$-$V$ characteristics along with numerical fits to the Eq. (1). Forward $I$-$V$ characteristics do not follow Eq. (1) and resulted in very high ideality factors ($n > 8$ for $V_F < 0.5$ V and even higher for larger biases). Such a large value of $n$ cannot be explained within the purview of diffusion or thermionic theory of junction diodes. Therefore, the form of Eq. (1) is not valid for analyzing the forward $I$-$V$ characteristics [4,8]. On the other hand the reverse $I$-$V$ characteristics showed comparatively smaller $n$ (< 2, for $V_R <$ 0.5 V) as shown in figure 2 (*b*). Low values of $n$ suggest the use of reverse $I$-$V$ characteristics to be a suitable choice for the analysis of device parameters.

The value of series resistance was found to be negligibly small compared to the heterojunction resistance for $|V| < 0.5$ V under both forward and reverse biases (figure 2). For $|V| < 0.5$ V, the $I$-$V$ characteristics are well described by Eq. (1) with $R_S=0$, [26]:

$$I = I_0 \exp(qV/nkT)\left[1-\exp(-qV/kT)\right] \quad (2)$$

Apart from barrier lowering, Eq. (2) takes into account the tunneling as well as carrier recombination effects in the depletion region of the junction [27]. Eq. (2) can also be used to analyze reverse bias characteristics, by rewriting it in the form:

$$\frac{I}{\left[1-\exp(-qV/kT)\right]} = I_0 \exp(qV/nkT) \quad (3)$$

The semi logarithmic plots of $I/\left[1-\exp(-qV/kT)\right]$ vs. $V$ for both forward and reverse biases are shown in figure 3. It demonstrates an excellent fit for the reverse bias. However, it is interesting to note that the forward $I$-$V$ characteristics do not lie on the same straight line as the reverse $I$-$V$ characteristics near zero bias, but bend downwards with large $n$ values. Thus the results in figure 3 further indicate that current transport mechanisms in forward and reverse conditions are different. The value of ideality factor $n$ obtained from the slopes of reverse $I$-$V$ in figure 3 increases with decreasing $T$ and lies between 1.30 to 1.93 for $T = 300 – 195$ K. $n$ becomes larger than 2 for $T < 195$ K. These values of $n$ suggest that the reverse current flows are mainly due to carrier generation-recombination in the depletion region. However, there are several other models which can describe the temperature dependence of ideality factor [26,28,29,30].

In order to explore the nature of temperature dependence and the greater-than-unity behavior of ideality factor, $nkT$ (the inverse slope of an $I$-$V$ curve) is plotted against $kT$ in figure 4. The data points are seen to lie in a straight line which when extrapolated, does not pass through the origin and has a slope different from unity. The considerable deviation of slope from unity negates the so called $T_0$ anomaly model for temperature dependent ideality factor, which is described by: $n=1+T_0/T$, where $T_0$ is a temperature independent constant [28,29]. Often the thermionic field emission (TFE) is invoked to explain the temperature dependence of $n$. If Eq. (3) is used to describe $I$-$V$ characteristics for TFE model, temperature dependent ideality factor in *reverse* bias becomes $nkT = E_{00}\{(qE_{00}/kT) - \tanh(qE_{00}/kT)\}^{-1}$, where $E_{00}$ is the temperature independent constant [26], and its temperature dependence is shown in figure 4. It is clear from figure 4 that TFE model is not applicable in our case. The presence of barrier height inhomogeneity and its effects on $I$-$V$ characteristics, or inhomogeneities related to metal-semiconductor junctions are not possible in our case of c-Si/PS heterojunction. More will be discussed on inhomogeneous barriers later on.

Now we invoke generation-recombination model to describe our results. According to this model, the temperature dependence of $n$ implies that the carrier generation-recombination involves defect states. The temperature dependence of $n$ may be modeled as [26]:

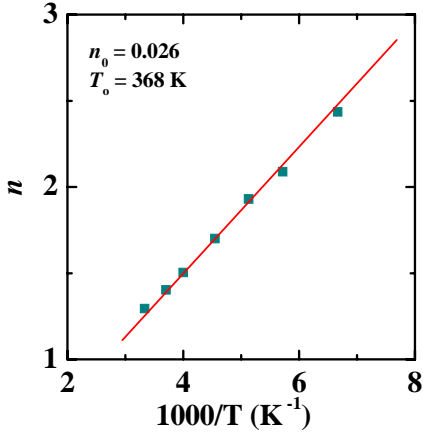

Figure 5. Temperature dependence of ideality factor $n$ as derived from the analysis shown in Figure 3.

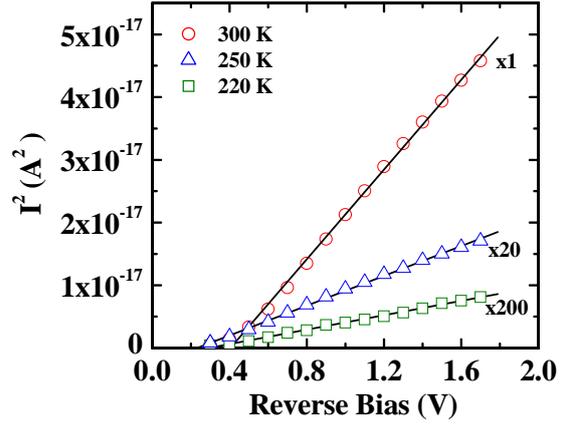

Figure 7. $I^2$ as a function of reverse bias voltage ($V$) at different temperatures. The values of $I^2$ are multiplied by a factor marked in the figure.

$$n = n_0 + T_0/T \qquad (4)$$

where $n_0$ and $T_0$ are constants which are independent of temperature and voltage. The variation of $n$ with inverse of temperature is demonstrated in figure 5, which fits to a straight line. The values of $n_0$ and $T_0$ were found to be ≈ 0.026 and ≈370 K, respectively. A very small value of $n_0$ implies the involvement of several defect levels in carrier generation-recombination processes. The large value of $T_0$ indicates that tunneling effects may also contribute to the reverse $I$-$V$ characteristics at lower temperatures, which is consistent with increasing value of $n$ below 195 K.

Further, the reverse saturation current $I_0$ can be written as [31]:

$$I_0 \propto \exp\left(-\frac{E_a}{nkT}\right) \qquad (5)$$

where $\phi_b$ is an energy barrier controlling the reverse currents. A semi logarithmic plot of $I_0$ against the reciprocal of the product of ideality factor $n$ and temperature $T$ is shown in figure 6, which demonstrates a reasonably good linear fit yielding $\phi_b$ ≈1.9 eV. The value of $\phi_b$ is too large to be the junction barrier height at $c$-Si/PS heterojunction. On the other hand, $\phi_b$ is quite close to the photoluminescence (PL) peak energy of ≈1.8 eV of PS layer measured on these samples [16], suggesting that the reverse current at zero bias in our sample is due to carrier generation-recombination in space charge region formed on the PS layer side of $c$-Si/PS heterojunction. Therefore, $\phi_b$ obtained from figures 3 and 6 represents the electrical bandgap of PS layer.

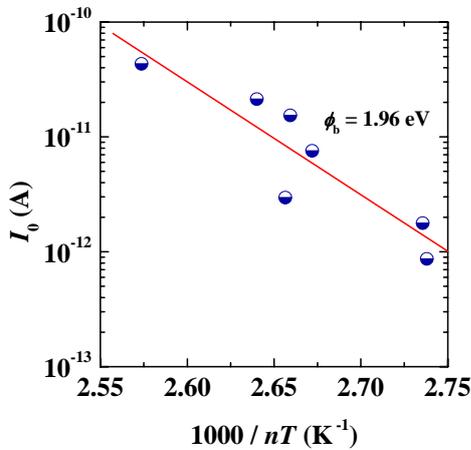

Figure 6. Arrhenius plot of reverse saturation current ($I_0$) obtained from Eq. 3.

Now, if carrier generation-recombination mechanism is dominant, the current should be proportional to $(V_d+V)^{1/2}$, where $V_d$ is the diffusion voltage at the junction [32]. This is indeed found to be the case as shown in figure 7 where $I^2$ is found to be linear with $V$ for reverse bias <2 V. The diffusion voltage $V_d$ was estimated to be 0.3-0.4 eV at different temperatures. Further, the information about bulk carrier transport in the PS layer may be obtained by analyzing the temperature dependence of currents in the reverse bias voltage range where $I^2$ is linear with $V$. The activation energy ($E_a$) from temperature dependent reverse currents at reverse bias of 1.2 V was found to be ≈ 0.21 eV (figure 8), which is similar to $E_a$ of PS layers obtained in planar configuration [19]. This result suggests that the depletion region is mainly formed in PS layer.

At higher reverse biases (> 5 V), the semi logarithmic plot between $I$ vs. $V^{1/2}$ exhibits nearly linear dependence as shown in figure 9. Such behavior is expected when the barrier lowering ($\Delta\phi$) is proportional to the electric field ($F$) present at Schottky junctions. This dependence on $F$ implies that the barrier lowering was dominated by mechanisms other than image-force lowering, $\Delta\phi_{image} = \sqrt{qF/4\pi\varepsilon_s}$, $\varepsilon_s$ being the dielectric constant of semiconductor [26,32]. However, the barrier lowering can be proportional to the electric filed $F$ when the Fermi level $E_F$ at the junction is pinned [33]. Therefore the ob-served $ln(I)$ vs. $V^{1/2}$ behavior suggests the pinning of Fermi level at the $c$-Si/PS interface. Other workers also have proposed such pinning and have attributed it to the presence of a large density of states at the $c$-Si/PS interface [4]. In this case, the barrier height ($\phi_b$) estimated from the temperature dependence of reverse saturation current $I_0$ [33] obtained in high reverse bias region would correspond to the band-discontinuity at the conduction band edges ($\Delta E_C$). In our study $\phi_b$ was found to be $\approx 0.13$ eV (figure 10).

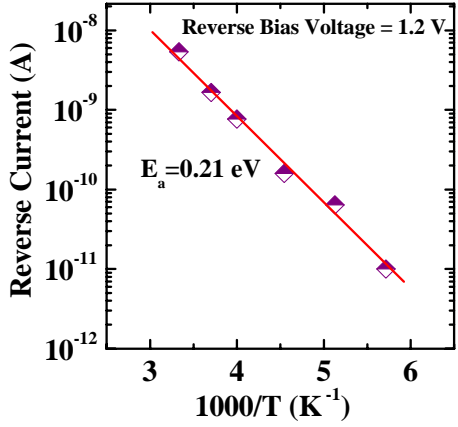

Figure 8. Arrhenius plot of reverse current at reverse bias of 1.2 V.

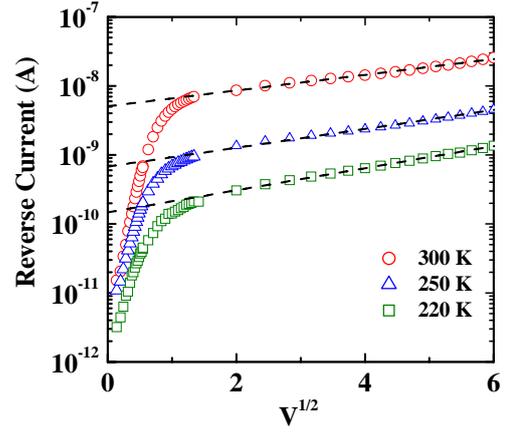

Figure 9. Semi logarithmic plots of reverse current $I$ vs. $V^{1/2}$ at different temperatures.

## 4. Discussion

For diffusion or thermionic emission of carriers, ideality factor $n$ is unity, whereas $n$ becomes equal to 2 for recombination-generation of carriers in depletion region of defect-free materials with Fermi level at the middle of the bandgap. However in practice, $n$ is found to vary from unity to very large value (>2). Furthermore, $n$ often varies with temperature. Several models have been proposed to explain the greater-than-unity and temperature dependent behavior of $n$ [26,28,29,30]. As discussed earlier, the $T_0$ anomaly and thermionic field emission could not explain the observed behavior of $n$ in our case. On the other hand, inhomogeneities in junction barrier heights (JBH) have been claimed to describe a very wide variety of $I$-$V$ characteristics except field emission [28,29]. However, the existence of inhomogeneous barriers must be explored very carefully.

Unlike in metal-semiconductor junctions, the possibility of existence of JBH inhomogeneity at our c-Si/PS heterojunction system is very low. The various factors which result in or are sources of such barrier height inhomogeneities are not present in our case. First, our heterojunction is sandwiched between c-Si and PS layer, both of which are crystalline, having inherently the same crystal orientation at the heterojunction interface. Hence this junction is free from all inhomogeneities related to polycrystalline metal surfaces such as grain boundaries, crystal orientations and intermetallic compound formation. Secondly, heterojunction in our study were made from a single piece of c-Si wafer, and it involved neither surface cleaning nor any external surface layer growth processes during c-Si/PS junction formation. By virtue of c-Si/PS heterojunction formation techniques described here, c-Si and PS layers form intimate contact. Thirdly, even if there is an existence of small JBH inhomogeneity due to the interface nonplanarity, its effects will be insignificant in lightly doped c-Si, and highly resistive PS layer [28,29]. Therefore, inhomogeneity in JBH [28,29] and / or distribution of JBH [30] cannot account for our results.

In our case, the ideality factor $n$ <2 and a low value of $n_0$ (figure 5) indicate that the recombination of electron-hole pairs occurs via a distribution of defect states. There are two dominant effects that decide the recombination current: the width of the depletion region ($W$) and the defect density in the bandgap ($N_t$). In $c$-Si, $N_t$ is very small and can be neglected. On the other hand, the defect levels in the bandgap are quasi-continuous in the porous

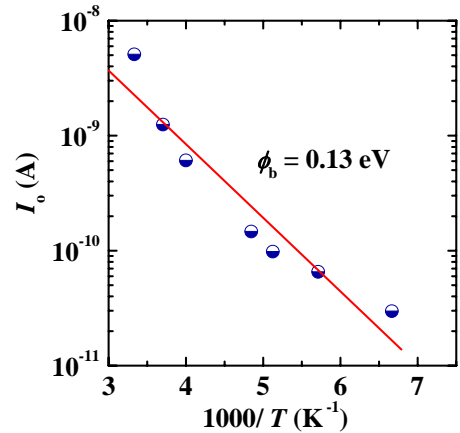

Figure 10. Arrhenius plot of $I_0$ obtained from high reverse bias region (figure 9).

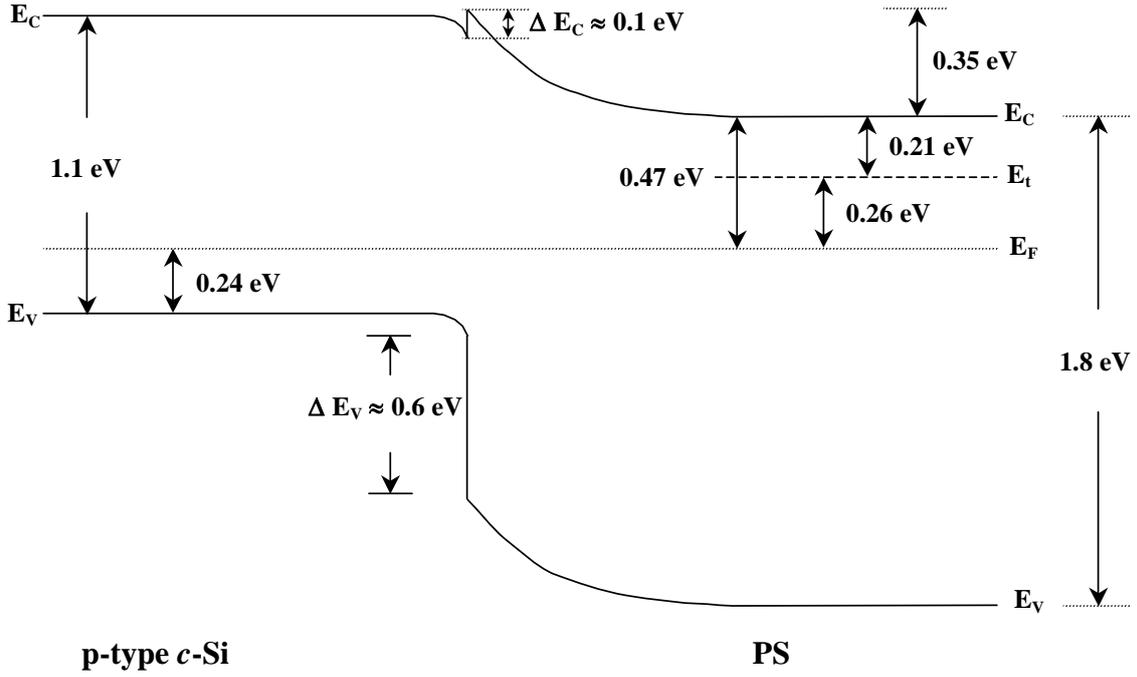

Figure 11. Schematic of deduced energy-band diagram of *c*-Si/ PS heterojunction.

Si like in the case of amorphous silicon. $N_t$ is reported/found to be $\approx 10^{19}$cm$^{-3}$eV$^{-1}$ in PS [34]. Furthermore, the activation energy $E_a$ ($\approx$ 0.21 eV) is much smaller compared to the half of the *c*-Si bandgap, whereas it is in fairly good agreement with the activation energy of PS layers [20,34] as well as with the energy of trap levels in PS layers [20,35]. Therefore, it is reasonable to expect that the recombination occur on the PS side of the depletion region. Further, a low value of $n$ is expected due to defect levels in the bandgap of PS layer [35]. The defects states in PS layer act as trap levels and induce recombination of carrier pairs. Recent photovoltage measurements also suggest that the depletion region at *c*-Si/PS heterojunction mainly forms on the PS side [36]. Thermovoltage measurements [20] on the PS layers prepared by a method similar to ours, show that majority carriers in PS layers are electrons. Scanning probe microscopy and scanning tunnelling spectroscopy studies have also indicated *n*-type behavior of PS layers prepared from *p*-type *c*-Si substrate [37].

Based on the conductivity of our *c*-Si substrate, we estimated the position of Fermi level at $\approx$ 0.24 eV above valence band edge ($E_V$) using standard procedure [32]. Further, the density of conduction electrons in semiconductor is given as:

$$n = N_C \exp\left[-(E_C - E_F)/kT\right] \quad (6)$$

However, at low temperatures where conduction electron density is much less than the trapped electron density contributing to conduction, the free electron density is controlled by the ionization of trapping states and is, similar to compensated materials, given by [38]:

$$n = \left(\frac{N_d}{2N_t}\right) N_C \exp\left[-(E_C - E_t)/kT\right] \quad (7)$$

Where $N_d$ and $N_t$ are the densities of trapped electrons in donor-like defect states and total electron trapping defect states at energy level of $E_t$ below $E_C$. The activation energy ($E_a$) of the conductivity thus becomes equal to ionization energy ($E_C - E_t$) of the trapped electrons. Therefore, we can write

$$\sigma = \sigma_0 \exp\left[-E_a/kT\right]$$

and

$$\sigma_0 = q\mu \left(\frac{N_d}{2N_t}\right) N_C \quad (8)$$

Using the geometrical parameters for our sample and value of $I_0$ (=2.17×10$^{-5}$A) obtained from temperature dependence of reverse current (figure 8), the pre-exponential factor $\sigma_0 = I_0 d / AV$ is estimated to be 2.8-5.5×10$^{-7}$($\Omega$cm)$^{-1}$. Moreover, the drift mobility $\mu$ in PS layer was measured to be $\approx 10^{-4}$–10$^{-3}$ cm$^{-2}$/Vs using optical and capacitance measurements [6]. Therefore, taking $\sigma_0 \approx 10^{-7}$ ($\Omega$cm)$^{-1}$, $\mu \approx 10^{-3}$ cm$^2$/Vs and $q \approx 10^{-19}$ C in Eq. (8), $N_d$ is found to be $\approx \left(\frac{2N_t}{N_C}\right) \times 10^{15}$ cm$^{-3}$.

The Fermi energy in PS may be calculated as follows.

$$n = \left(\frac{N_d}{2N_t}\right) N_C \exp\left[-(E_c - E_t)/kT\right] =$$

$$N_C \exp\left[-(E_C - E_F)/kT\right] \tag{9}$$

or
$$E_C - E_F = E_C - E_t + kT \ln\left(\frac{2N_t}{N_d}\right) \tag{10}$$

Now, with $E_C - E_t = E_a = 0.21\,\text{eV}$ (figure 8) and $N_t$ ($<N_C$) $\approx 10^{19}$ cm$^{-3}$, the value of $E_C - E_F$ for porous Si comes out to be $\approx 0.47$ eV at room temperature.

Further, if we consider the average peak PL energy as a measure of average bandgap of our PS sample ($\approx 1.8$ eV), the valence band discontinuity may be estimated as:

$$\Delta E_V = {}^{PS}E_g - {}^{Si}E_g - \Delta E_C \approx 1.8\text{-}1.1\text{-}0.1 \text{ eV} \approx 0.6 \text{ eV}$$

Here, this value compares well with the experimental value of 0.6 eV obtained using photoconductivity measurements [39,40]. Based on the above analysis, we propose the band gap line-up at the $c$-Si/PS heterojunction as shown in figure 11. Needless to say, the information regarding the band edge discontinuities of PS is very important for the successful use of this material in optoelectronic devices. We can see in figure 11 that one can resort to the basic physics of semiconductor junction properties to obtain band discontinuities in c-Si/PS heterojunctions in a very simple manner.

## 5. Conclusions

We have prepared light emitting nanocrystalline porous Si layers by electrochemical etching of $c$-Si wafers and characterized the $c$-Si/PS heterojunctions using temperature dependence of dark $I$-$V$ characteristics. In particular, reverse bias $I$-$V$ characteristics are found to yield valuable information on the current transport across the $c$-Si/PS junction. The current transport mechanism in the reverse bias condition is mainly dominated by the carrier generation-recombination in the depletion region formed on PS side. At higher reverse biases, the reverse current transport is governed by the barrier lowering effects. It is found to behave like a Schottky junction with Fermi level pinned to the defect energy levels at the $c$-Si/PS interface. The conduction band offset is found to be $\approx 0.1$ eV. Based on the detailed analysis of $I$-$V$ data the energy band diagram of the $c$-Si/PS heterojunction has been presented. Our study provides an easy and useful alternative method of determination of band edge discontinuities in multilayer structures using PS layers.

## Acknowledgments

One of the authors (MNI) acknowledges the support and permission given by Indian Space Research Organization, Ahmedabad to publish this work.